\documentclass[%
 reprint,
 superscriptaddress,
preprintnumbers,
nofootinbib,
 amsmath,amssymb,
 aps,
]{revtex4-1}

\usepackage{tikz}
\usepackage{graphicx}
\usepackage{dcolumn}
\usepackage{bm}
\usepackage{hyperref}
\usepackage{color}
\usepackage{multirow, array}

\hypersetup{
    colorlinks = true
}
\usepackage{aas_macros}

\newcommand{\be}{\begin{equation}}
\newcommand{\ee}{\end{equation}}
\newcommand{\bea}{\begin{eqnarray}}
\newcommand{\eea}{\end{eqnarray}}
\newcommand{\nn}{\nonumber}

\newcommand{\vpt}{\mbox{\sf VPT}}
\newcommand{\fastvpt}{\mbox{\sf fVPT}}

\begin{document}

\title{Vlasov Perturbation Theory applied to $\Lambda$CDM}

\author{Mathias Garny}
\email{mathias.garny@tum.de}
\affiliation{Physik Department T31, Technische Universit\"at M\"unchen, James-Franck-Stra\ss{}e 1, D-85748 Garching, Germany
}%

\author{Dominik Laxhuber}
\email{dominik.laxhuber@tum.de}
\affiliation{Physik Department T31, Technische Universit\"at M\"unchen, James-Franck-Stra\ss{}e 1, D-85748 Garching, Germany
}%

\author{Rom\'an Scoccimarro}
\email{rs123@nyu.edu}
\affiliation{
 Center for Cosmology and Particle Physics, Department of Physics, New York University, NY 10003, New York, USA
}%

\date{05.05.2025}

\begin{abstract} 

We apply the framework of Vlasov Perturbation Theory (\vpt{}) to the two-loop matter power spectrum within $\Lambda$CDM cosmologies.  The main difference to Standard Perturbation Theory (SPT) arises from taking the velocity dispersion tensor into account, and the resulting screening of the backreaction of UV modes renders loop integrals cutoff-independent. \vpt{} is informed about non-perturbative small-scale dynamics via the average value of the dispersion generated by shell-crossing, which impacts the evolution of perturbations on weakly non-linear scales. When using an average dispersion from halo models, the \vpt{} power spectrum agrees with the one from the simulation, up to differences from missing three-loop contributions. Alternatively, treating the average dispersion as free parameter we find a remarkably stable prediction of the matter power spectrum from collisionless dynamics at percent level for a wide range of the dispersion scale.
We quantify the impact of truncating the Vlasov hierarchy for the cumulants of the phase-space distribution function, finding that the two-loop matter power spectrum is robust to neglecting third and higher cumulants. Finally, we introduce and validate a simplified fast scheme \fastvpt{} that can be easily incorporated into existing codes and is as numerically  efficient as SPT.

\end{abstract}

\maketitle

%===========================================================
\section{Introduction}
\label{sec:introduction}
%===========================================================

\begin{figure*}[t]
  \begin{center}
  \includegraphics[width=0.6\textwidth]{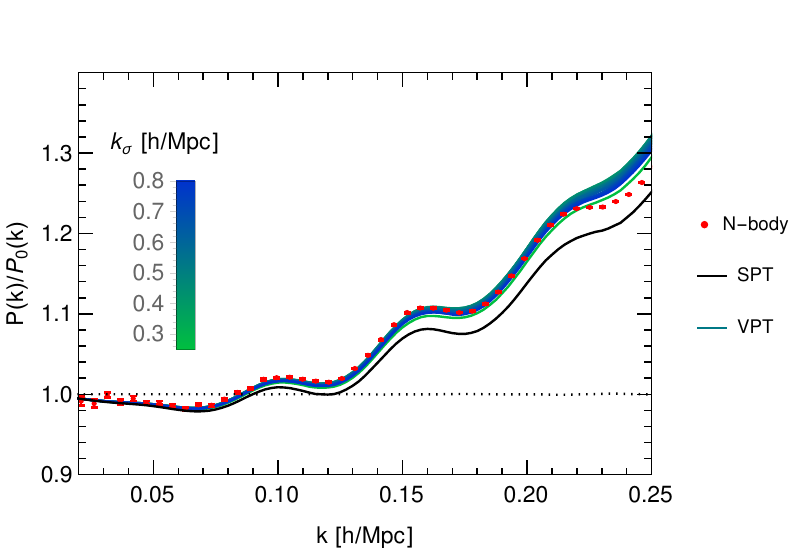}
  \end{center}
  \caption{\label{fig:Pdd2L}
  Matter power spectrum $P(k)$ at redshift $z=0.34$ in \vpt{}, compared to $N$-body simulation results.
  The various coloured lines for \vpt{} show the (small) dependence on the background dispersion scale $k_\sigma$ which is varied by more than a factor of three, as labeled.
  We additionally find the \vpt{} result to be highly insensitive to the truncation order of the Vlasov cumulant hierarchy,
  the average value of higher cumulants, as well as the UV cutoff, making it a genuine prediction for collisionless matter clustering.
  For illustration we also show the SPT result, which could only be brought in agreement with $N$-body results
  when adding free fitting parameters within the EFT approach. Both \vpt{} and SPT are evaluated at two-loop order. Residual differences to the $N$-body result can be associated to missing
  three- and higher loop corrections. We also highlight the non-monotonic dependence of $P(k)$ on $k_\sigma$, see Secs.~\ref{sec:ksigmadependence} and\,\ref{sec:ksigmadependencefvpt} for details.
  }
\end{figure*}

Ongoing stage-IV galaxy surveys such as DESI~\cite{DESI:2016fyo} and Euclid~\cite{Euclid:2024yrr} call for pushing theoretical predictions of structure formation to
a new level, covering in particular the weakly non-linear regime for which information on the underlying cosmological model can be retrieved using semi-analytical perturbative methods, based on first-principle calculations of galaxy clustering that respect the underlying symmetries.

Within the standard $\Lambda$CDM model of cosmology, the formation of 
structure in the universe is driven at large scales by gravity acting on collisionless cold dark matter working against the expansion of the universe, which at late times accelerates due to a cosmological constant. This dynamics is described by the Vlasov (or collisionless Boltzmann) equation  coupled to Poissonian gravity for the most relevant scales (well below the Hubble radius). While this dynamical system underlies the $N$-body method, perturbative techniques such as Standard Perturbation Theory (SPT) (see~\cite{Bernardeau:2001qr} for a review) are based instead on a fluid description, 
leading to well-known limitations such as a spurious sensitivity to backreaction from UV modes, see \emph{e.g.}~\cite{peebles1980large,BerTarNis1401,BlaGarKon1309,NisBerTar1611,NisBerTar1712}. The fluid approximation ultimately restricts either the accuracy or, when
including correction terms with free parameters within the effective field theory (EFT) approach~\cite{BauNicSen1207}, the predictivity.

This motivates a perturbative treatment that is directly derived from the Vlasov-Poisson system, on equal footing with $N$-body simulations. This is precisely the objective of 
 Vlasov Perturbation Theory (\vpt{})~\cite{cumPT,cumPT2,Garny:2025ovs}, see also~\cite{Pueblas:2008uv,McD1104,Col1501,Aviles_2016,Uhlemann:2018olp,McDVla1801,SagTarCol1812,ErsFlo1906,Erschfeld:2023aqr,Nascimento:2024hle} for related work along these lines. The novel feature of Vlasov-Poisson dynamics, as opposed to a fluid, is that orbit crossing of collisionless particles generates dispersion and higher cumulants of the distribution function. This means that the velocity dispersion tensor develops a non-zero spatial average, or expectation value, which in \vpt{} defines a (time-dependent) dispersion scale $k_\sigma$ comparable to the nonlinear scale. Fourier modes that cross such scale ($k>k_\sigma$) stop growing, an effect already present in  \vpt{} at the linear level, leading to non-trivial linear kernels for perturbations. At the non-linear level, this gives rise to a large suppression of the mode-coupling between large-scale and small-scale modes $q>k_\sigma$, capturing the physically expected screening of the backreaction of UV modes within perturbation theory, absent in SPT and the cause of its demise. Furthermore, since the dispersion scale is close to the nonlinear scale, loops in \vpt{} get most of their contribution from modes within perturbative control.

In previous work~\cite{cumPT,cumPT2,Garny:2025ovs}, we applied \vpt{} to scale-free cosmologies with blue spectral indices, for which effects of orbit crossing and velocity dispersion are enhanced and SPT becomes UV divergent. This provided a good stress test for \vpt{}, and led to a number of important results: 
\begin{enumerate}

\item[i)] \vpt{} captures UV screening and leads to finite results at any loop order and for any spectral
index consistent with hierarchical clustering; 
\item[ii)] all qualitative features of \vpt{} are captured already when including the velocity dispersion tensor (second cumulant) alone, which gives a significant difference from SPT no matter how small the dispersion is (\emph{i.e.}~for spectral indices $n_s\geq-1$ it gives finite answers to loop corrections, versus divergent answers in SPT). On the other hand, the impact of truncating
the Vlasov hierarchy at higher order in the cumulant expansion can be systematically computed and was found to be small for density and velocity power spectra, its main effect being making the screening of UV modes even more efficient; 
\item[iii)] both scalar
and vector modes (such as vorticity) are relevant while tensor and higher-helicity modes are highly suppressed; average values of fourth and higher cumulants are quantitatively by far
subdominant compared to the average of the second cumulant (related to $k_\sigma$), and are furthermore constrained by universal (independent of spectral index) stability
conditions; 
\item[iv)] \vpt{} results for density and velocity spectra as well as the bispectrum agree well with $N$-body simulations when taking the dispersion scale $k_\sigma$ as a single non-perturbative input parameter.
\end{enumerate}
In this work we extend these results and apply \vpt{} to a realistic $\Lambda$CDM cosmology. We focus on the matter density power spectrum in real space, and present \vpt{} results
up to two-loop order. It is well-known that two-loop SPT is highly dependent on the UV cutoff as a consequence of using a fluid description. We analyze in detail how \vpt{} overcomes
limitations of SPT. We quantitatively assess the role of the dispersion scale $k_\sigma$, demonstrate robustness with respect to truncating the Vlasov hierarchy, and verify the
physically expected UV insensitivity. When taking $k_\sigma$ as non-perturbative input, we arrive at a prediction for collisionless matter clustering that can be unambigously compared to $N$-body results.

As a first possibility, we extract the dispersion scale from the simulation itself, leading to consistent results. Alternatively, we allow for variations in $k_\sigma$ and find a remarkable
insensitivity of the two-loop matter power spectrum to the precise value of this scale. This result is summarized in Fig.~\ref{fig:Pdd2L}, where the \vpt{} prediction obtained for a large range of
choices for $k_\sigma$, spanning over a factor of three, is shown. Nevertheless, it varies only at the one percent level on weakly non-linear scales $k\lesssim 0.2h/$Mpc. Together with our findings regarding
truncation-independence (at sub-percent level) this \vpt{} result can thus be viewed as a genuine prediction for collisionless matter clustering, up to the missing three- and higher loop contributions.

After a review of the \vpt{} setup in Sec.\,\ref{sec:setup}, we present and discuss our results for the matter power spectrum up to two-loop level in Sec.\,\ref{sec:pk}.
Point ii) above, that all qualitative effects of \vpt{} are present with dispersion alone, added to the fact that its effects already appear in linear theory, raises the question of whether one can approximate \vpt{} by a simplified scheme that is much faster to compute. 
In Sec.\,\ref{sec:vpt}, we propose and validate such approximate scheme that we refer to as \fastvpt{}. It is numerically as efficient as SPT and can be easily included in SPT-based implementations.
We use \fastvpt{} to analyze the $k_\sigma$-dependence in detail and comment also on relations to the EFT approach, before
concluding in Sec.\,\ref{sec:conclusions}. The Appendix demonstrates independence of our results on the UV cutoff.

For illustration, throughout this work we compare our results to $N$-body data from {\tt LasDamas}-{\it Oriana}~\cite{McBBerSco0901,SinBerMcB1807,Crocce:2012} runs using the {\tt Gadget2} code~\cite{Spr0512} with {\tt 2LPT}
initial conditions~\cite{Scoccimarro:1998b,CroPueSco0611} (box size $L=2400$Mpc$/h$, $N=1280^3$), with cosmological parameters $\Omega_m=0.25, \Omega_b=0.04, h=0.7, \sigma_8=0.8, n_s=1$.

%===========================================================
\section{Vlasov Perturbation Theory setup}
\label{sec:setup}
%===========================================================

Clustering of collisionless matter is governed by the Vlasov equation,
\be
  0 = \frac{df}{d\tau}=\frac{\partial f}{\partial\tau}+\frac{p_i}{a}\frac{\partial f}{\partial x_i} -a(\nabla_i\Phi)\frac{\partial f}{\partial p_i} \,,
\ee
for the phase-space distribution function $f(\tau,{\bm x},{\bm p})$ along with the Poisson equation $\nabla^2\Phi(\tau,{\bm x})=\frac32{\cal H}^2\Omega_m\,\delta(\tau,{\bm x})$
for the gravitational potential $\Phi$ sourced by the density contrast $1+\delta=\int d^3p\,f(\tau,{\bm x},{\bm p})$.
Here ${\cal H}=d\ln a/d\ln\tau$ with $a$ the  scale factor measuring the expansion of the universe, conformal time $\tau$ and time-dependent matter density parameter $\Omega_m$.

\vpt{}~\cite{cumPT,cumPT2,Garny:2025ovs} is a systematic framework for solving the Vlasov-Poisson equations for the perturbations of the distribution function.
In a first step, the Vlasov equation is rewritten into a coupled set of equations for its cumulants ${\cal C}_{i_1i_2\cdots i_c}(\tau,{\bm x})=\partial^n{\cal C}/\partial_{l_{i_1}}\cdots\partial_{l_{i_c}}|_{{\bm l}=0}$
derived from the generating function
\be
  e^{{\cal C}(\tau,{\bm x},{\bm l})} \equiv \int d^3p \, e^{{\bm l}\cdot{\bm p}/a}\, f(\tau,{\bm x},{\bm p})\,.
\ee
The equations for the zeroth cumulant (related to $\delta(\tau,{\bm x})$), the first cumulant $v_i(\tau,{\bm x})\equiv {\cal C}_i$ ({\it i.e.} the peculiar velocity field), the second cumulant $\sigma_{ij}(\tau,{\bm x})={\cal C}_{ij}$ ({\it i.e.} the velocity dispersion tensor) and the third one read
\bea\label{eq:eom}
  \partial_\tau\delta + \nabla_i v_i &=& -\nabla_i(\delta v_i) \,,\nn\\
  \partial_\tau v_i+{\cal H}v_i+\nabla_i\Phi + \nabla_j\sigma_{ij}&=& -v_j\nabla_j v_i-\sigma_{ij}\nabla_jA\,,\nn\\
   \partial_\tau\sigma_{ij} +2{\cal H}\sigma_{ij} +\nabla_k{\cal C}_{ijk} &=&
   -v_k\nabla_k\sigma_{ij}-\sigma_{jk}\nabla_k v_i \nn\\
  && {} -\sigma_{ik}\nabla_k v_j -{\cal C}_{ijk}\nabla_kA\,,\nn\\
   \partial_\tau{\cal C}_{ijk} +3{\cal H}{\cal C}_{ijk} +\nabla_m{\cal C}_{ijkm} &=& -v_m\nabla_m{\cal C}_{ijk} - (\sigma_{km}\nabla_m\sigma_{ij}\nn\\
   && {} +{\cal C}_{jkm}\nabla_m v_i +2\,\text{perm.})\nn\\
   && {} - {\cal C}_{ijkm}\nabla_m A\,,
\eea
where $A\equiv \ln(1+\delta)$ with $\partial_\tau A+\nabla_i v_i=-v_i\nabla_i A$. The equation for the $c$th cumulant depends on those up to $c+1$, such
that the coupled hierarchy needs to be truncated at some order $c\leq c_\text{max}$. 
The truncation $c_\text{max}=1$ corresponds to the pressure-less perfect  fluid approximation underlying SPT, while \vpt{} assumes $c_\text{max}\geq 2$.
The dependence on the truncation order has been investigated in~\cite{Garny:2025ovs}, finding that all essential features of \vpt{} are present
already for $c_\text{max}=2$, {\it i.e.} when including the velocity dispersion tensor as a dynamical variable. Ref.~\cite{Garny:2025ovs} also provided a setting for systematically
solving the hierarchy up to -- in principle -- arbitrarily large $c_\text{max}$, finding a small quantitative impact on the matter power spectrum from cumulants beyond second order
in scale-free models. In Sec.\,\ref{sec:pk} we extend this analysis to $\Lambda$CDM.

\begin{table*}[t]
  \centering
  \caption{Summary of \vpt{} approximation schemes used in this work. All of them take the average dispersion $\epsilon(z)=k_\sigma(z)^{-2}$ into account. For perturbation modes, we consider the second cumulant truncation $c_\text{max}=2$ as fiducial choice, as well
  as $c_\text{max}=3$. For $c_\text{max}=3$ we also take into account the average value $\bar{\cal E}_4$ of the fourth cumulant, which enters in the evolution equations for third-cumulant perturbations. In both cases we include scalar and vector modes, but omit tensor and higher modes (their impact on density kernels is highly suppressed, see~\cite{Garny:2025ovs}).  The notation for the spherical (cartesian) basis for cumulant perturbations follows~\cite{Garny:2025ovs} (\cite{cumPT}).  Note that for technical reasons we further include the log-density $\delta C_{000}=A-\langle A\rangle$ in addition to $\delta$, see~\cite{cumPT}. In this work, we additionally introduce a simplified scheme \fastvpt{}, which captures the dominant effects from velocity dispersion on the density field (including the average dispersion $\epsilon(z)$ and the impact of scalar perturbation modes $g$ and $\delta\epsilon$ on linear density and velocity kernels; see Sec.\,\ref{sec:vpt} for details) but
  is computationally as fast as SPT. For a detailed study on the truncation-dependence within full \vpt{}, including perturbations of fourth and higher cumulants ($c_\text{max}=4,5,6$) as well as tensor and higher-helicity modes, we refer
  to~\cite{Garny:2025ovs}.}
  \begin{ruledtabular}
  \begin{tabular}{l|c|c|c|c|c|c|c|c|c|c|c|c} 
  cumulant order \rule[-1.5mm]{0mm}{4mm} & 0 & \multicolumn{2}{l|}{1 \ \ (velocity $u_i$)} & \multicolumn{4}{c|}{2 \ \ (vel. dispersion $\epsilon_{ij}$)} & \multicolumn{4}{c|}{3 \ \ (${\cal C}_{ijk}$)} & 4\\ \hline
  type  \rule[-1mm]{0mm}{4mm}          & pert. & \multicolumn{2}{l|}{perturbation} & \multicolumn{3}{c|}{perturbation $\delta\epsilon_{ij}$} & av. & \multicolumn{4}{c|}{perturbation} & av.\\ \hline
  pert. type  \rule[-1mm]{0mm}{4mm}          & scalar & scalar & vector & scalar & vector & tensor & - & scalar & vector & tensor & $|m|=3$ & -\\ \hline
  number of dof \rule[-1mm]{0mm}{4mm}  & 1 & 1 & 2 & 2 & 2 &  2 & - & 2 & 4 & 2 & 2 & -\\ \hline
  spherical basis~\cite{Garny:2025ovs} \rule[-1mm]{0mm}{4mm} & $\delta$ & $\delta C_{100}$ & $\delta C_{1,\pm1,0}$  & $ \delta C_{200}$  & $\delta C_{2,\pm 1,0}$ & $\delta C_{2,\pm 2,0}$ &$\epsilon(z)$& $\delta C_{300}$  & $\delta C_{3,\pm 1,0}$ & $\delta C_{3,\pm 2,0}$ & $\delta C_{3,\pm 3,0}$ & ${\cal E}_4$\\ 
  (used here)  \rule[-1mm]{0mm}{4mm} &  & $=\theta/3$ &   & $\delta C_{002}$  &  & &  & $\delta C_{102}$  & $\delta C_{1,\pm 1,2}$ &&\\ \hline
  cartesian basis~\cite{cumPT} \rule[-1mm]{0mm}{4mm} & $\delta$ & $\theta=\nabla_iu_i$ & $\varepsilon_{ijk}\nabla_ju_k$ & $g,\delta\epsilon$ & $\nu_i$ & $t_{ij}$ &$\epsilon(z)$& $\pi,\chi$ & - &  - & - & ${\cal E}_4$ \\ \hline
  SPT \rule[-1mm]{0mm}{4mm} & $\checkmark$ & $\checkmark$ &   &  &  & & & & & &\\
  \vpt{} $c_\text{max}=2$ \rule[-1mm]{0mm}{4mm} & $\checkmark$ & $\checkmark$ & $\checkmark$ & $\checkmark$  & $\checkmark$ & & $\checkmark$  & & & &\\
  \vpt{} $c_\text{max}=3$ \rule[-1mm]{0mm}{4mm} & $\checkmark$ & $\checkmark$ & $\checkmark$ & $\checkmark$  & $\checkmark$ & & $\checkmark$  & $\checkmark$ & $\checkmark$ & & & $\checkmark$\\
  \fastvpt{} $c_\text{max}=2$ \rule[-1mm]{0mm}{4mm} & $\checkmark$ & $\checkmark$ &  & ($\checkmark$)  &  & & $\checkmark$  &  &  & & & \\
  \end{tabular}
  \end{ruledtabular}
  \label{tab:modes}
\end{table*}

For convenience we introduce the rescaled variables
\be
  u_i\equiv\frac{v_i}{-f{\cal H}},\ \epsilon_{ij}\equiv \frac{\sigma_{ij}}{(f{\cal H})^2}\,,
\ee
and similarly for higher cumulants. Here $f=d\ln D/d\ln a$ is the growth rate, and $D(z)$ the usual growth function.

A key feature of \vpt{} is the split of cumulants into perturbation and average values.
The latter are generated by small-scale dynamics via shell crossing, and can therefore be
treated as physical, non-perturbative input in \vpt{}, sourcing in turn (small) perturbations of
second and higher cumulants on weakly non-linear scales, affecting the evolution of the density field.
The lowest cumulant that can possess an average value allowed by isotropy and conservation laws is the dispersion tensor\footnote{the zeroth cumulant, $A= \ln(1+\delta)$, has an expectation value but only gradients of it appear in the equations of motion, see Eq.~(\ref{eq:eom}).},
\be
  \langle\epsilon_{ij}(\tau,{\bm x})\rangle = \epsilon(z)\ \delta_{ij}^K\,,
\ee
with Kronecker symbol $\delta^K$ and $\epsilon(z)=\langle\sigma_{ii}\rangle/(3(f{\cal H})^2)$. The average dispersion $\epsilon(z)$ introduces a new scale,
\be
  k_\sigma(z) \equiv \epsilon(z)^{-1/2}\,,
\ee
with value of the order of the non-linear scale~\cite{cumPT2}. We also use the notation 
\be
  k_\sigma\equiv k_\sigma(0)\,.
\ee
While we discuss the choice and impact of $k_\sigma(z)$ in detail below, we stress already that
the appearance of this scale leads to screening of UV modes above $k_\sigma$ within \vpt{}, resulting in perturbative predictions
that are dominated by modes under theoretical control. We emphasize that this feature is not put by hand, but
a consequence of Vlasov-Poisson dynamics, in agreement with expectations~\cite{peebles1980large} and simulation results~\cite{NisBerTar1611,NisBerTar1712}. The next cumulant
with an average is at order four, given by ${\cal E}_4(z)\equiv \langle{\cal C}_{iijj}\rangle/(5(f{\cal H})^4)$. We
parameterize it by the kurtosis, or dimensionless ratio $\bar{\cal E}_4\equiv {\cal E}_4/\epsilon^2$.

We then define perturbations $\delta {\cal C}_{ij\cdots}\equiv {\cal C}_{ij\cdots}-\langle {\cal C}_{ij\cdots}\rangle$ around the average,
and insert this split into Eq.\,\eqref{eq:eom} to obtain equations of motion for $\delta {\cal C}_{ij\cdots}$. Furthermore, the cumulant
perturbations are decomposed into scalar, vector and tensor modes, as well as higher-helicity modes for cumulant order three or more, see~\cite{Garny:2025ovs} for details.
For example, the velocity field $u_i$ contains one scalar mode (related to $\theta=\nabla_i u_i$) and two vector modes related to vorticity $\nabla\times{\bf u}$.
The perturbations of the dispersion tensor $\delta\epsilon_{ij}=\epsilon_{ij}-\epsilon \delta_{ij}^K$ contain two scalar, two vector and two
tensor degrees of freedom. All perturbation modes can be collected into a large vector $\psi_a=(\delta,A,u_i,\delta\epsilon_{ij},\dots)$, with a generalized
index $a$ labelling all individual components. The modes included in this work are summarized in Tab.\,\ref{tab:modes}.

In Fourier space, the equation of motion for the cumulant perturbations takes the form
\bea\label{eq:psieom}
  &&\frac{\partial}{\partial\ln D(z)}\psi_a(z,{\bm k}) + \Omega_{ab}(z,k)\psi_b(z,{\bm k}) \nn\\
  &=& \int d^3p d^3q \, \delta_D^{(3)}({\bm k}-{\bm p}-{\bm q})\gamma_{abc}({\bm p},{\bm q})\psi_b(z,{\bm p})\psi_c(z,{\bm q})\,.\nn\\
\eea
The linear evolution matrix $\Omega_{ab}(z,k)$ depends on time and scale, via $\epsilon(z)k^2$, and $\bar{\cal E}_4$ (or in general even higher average
cumulants). $\bar{\cal E}_4$ appears when including {\it third} cumulant perturbations or higher. Note that we use a truncation scheme where we account for $\bar{\cal E}_4$ even when
truncating the cumulant {\it perturbations} at third order. Non-linear mode-coupling is described by the vertices $\gamma_{abc}$, see~\cite{cumPT} (\cite{Garny:2025ovs}) for explicit
expressions in cartesian (spherical harmonic) bases for the cumulant perturbations, with the latter including the most general case at arbitrary cumulant order. The vertices mix
modes of different type, {\it e.g.} scalar and vector modes, and capture vorticity generation and backreaction~\cite{cumPT2}.
The perturbations of each degree of freedom $\psi_a$ can be expanded in terms of the initial density contrast $\delta_{{\bm k}0}=\delta_0({\bm k})$,
\bea\label{eq:Fna}
  \psi_a(z,{\bm k}) &=& \sum_n \int d^3k_1\cdots d^3k_n\,\delta_D^{(3)}({\bm k}-{\bm k}_1-\cdots -{\bm k}_n)\nn\\
  && {} D(z)^n\, F_{n,a}({\bm k}_1,\dots,{\bm k}_n;z)\delta_{{\bm k}_10}\cdots\delta_{{\bm k}_n0}\,,
\eea
with non-linear kernels $F_{n,a}({\bm k}_1,\dots,{\bm k}_n;z)$ for all modes labelled by the generalized index $a$ and orders $n\geq 1$. In particular, the kernels $F_n\equiv F_{n,a=\delta}$
generalize those from SPT to \vpt{}. Inserting Eq.\,\eqref{eq:Fna} into Eq.\,\eqref{eq:psieom} yields evolution equation for the kernels, that can be solved recursively.
The matter power spectrum at linear, one-loop and two-loop order can be computed in \vpt{} analogously as in SPT, by replacing the SPT by \vpt{} $F_n$ kernels, and taking into
account that also the $F_1$ kernel is non-trivial in \vpt{},
\bea\label{eq:P1L2L}
  P_\text{lin}(k,z) &=& D(z)^2F_1(k,z)^2P_0(k)\,,\nn\\
  P_{1L}(k,z) &=& D(z)^4\int d^3p\,\big[6F_1(k,z)F_3({\bm k},{\bm p},-{\bm p},z)P_0(k)\nn\\
  && {} + 2F_2({\bm k}-{\bm p},{\bm p},z)^2P_0(|{\bm k}-{\bm p}|)\big]P_0(p)\,,\nn\\
  P_{2L}(k,z) &=& D(z)^6\int d^3p\,d^3q\,\big[\nn\\
  && \big( 30F_1(k,z)F_5({\bm k},{\bm p},-{\bm p},{\bm q},-{\bm q},z) \nn\\
  && {} + 9F_3({\bm k},{\bm p},-{\bm p},z)F_3({\bm k},{\bm q},-{\bm q},z)\big)P_0(k)\nn\\
  && {} + 6F_3({\bm k}-{\bm q}-{\bm q},{\bm p},{\bm q},z)^2P_0(|{\bm k}-{\bm p}-{\bm q}|)\nn\\
  && {} + 24F_2({\bm k}-{\bm p},{\bm p},z)F_4({\bm k}-{\bm p},{\bm p},{\bm q},-{\bm q},z)\nn\\
  && {} \times P_0(|{\bm k}-{\bm p}|)\big]P_0(p)P_0(q)\,.
\eea
Here $P_0(k)$ is the usual linear input spectrum.
For given $\epsilon(z)$ and $\bar{\cal E}_4$, we compute the kernels entering the loop integrand numerically, see Sec.\,IV\,A in~\cite{Garny:2025ovs} for details, and compute the
loop integrals as described in~\cite{Blas:2013bpa,BlaGarKon1309}.
The \vpt{} kernels are suppressed compared to SPT when one or more wavenumber arguments exceed $k_\sigma(z)$~\cite{cumPT2}, capturing the physically expected effect of UV screening mentioned above.

\medskip

\begin{figure}[t]
  \begin{center}
  \includegraphics[width=\columnwidth]{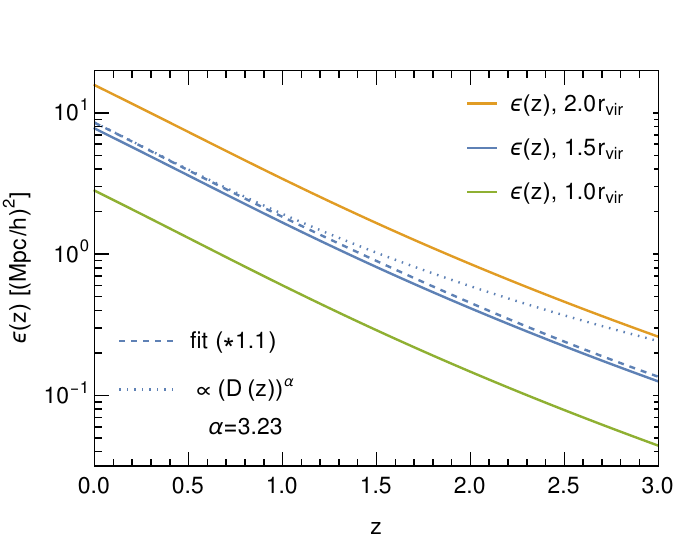}
  \end{center}
  \caption{\label{fig:epsilon}
  Average velocity dispersion $\epsilon(z)=\frac13\langle\sigma_{ii}\rangle/(f{\cal H})^2$
  estimated from halo profiles, associating the contribution from halos with their dispersion up to $1.0, 1.5, 2.0$ times the virial radius.
  We also show a broken power-law fit (dashed) and a single power-law approximation $\epsilon(z)\propto D(z)^\alpha= e^{\alpha\eta}$ for low $z$ (dotted), both
  offset by a factor $1.1$ for better readability. The average dispersion determines the scale $k_\sigma=\epsilon^{-1/2}$ which enters \vpt{} as a non-perturbative input.
  }
\end{figure}

Within \vpt{}, the average dispersion $\epsilon(z)$ is taken as non-perturbative input, and we consider two distinct parameterizations in this work.
As we will see, \vpt{} predictions are only mildly dependent on its choice.
First, we consider a simple power-law parameterization in terms of the growth function $D(z)$,
\be\label{eq:powerlaw}
  \epsilon(z)=\epsilon\,D(z)^\alpha\,,
\ee
with two parameters $\alpha$ and $\epsilon\equiv k_\sigma^{-2}$.
Second, we use a determination based on a halo model approach~\cite{cumPT2} informed by $N$-body simulation results,
\be\label{eq:halo}
  \epsilon(z) = \int \bar\epsilon_h(m,z)f(m,z)dm\,,
\ee
where $f(m,z)=\bar\rho/m\, dn/dm$ is the fraction of mass in halos of mass $m$, being related to the halo mass function $dn/dm$,
and $\bar\epsilon_h(m,z)$ is the dispersion averaged over a single halo. 
For $dn/dm$ we use a simple parameterization~\cite{SheTor9909,DesGioAng1603} with parameters fit to $N$-body results, while $\bar\epsilon_h(m,z)$
is computed assuming a Navarro-Frenk-White (NFW) halo, see Eq.~(118) in~\cite{cumPT}. This approach assumes all mass (and dispersion) is in dark matter halos. To account for dispersion outside halos, we assign to each halo a dispersion that sums up contributions up to $r\leq (1.5,2.0)r_\text{vir}$, in addition to the canonical choice of one virial radius (also calculated) that can be considered as a lower limit. Given that dispersion is described reasonably well also somewhat outside of the virial radius for NFW halos~\cite{ColLac9607}, we consider $1.5\, r_\text{vir}$ as fiducial choice while $2.0 \,r_\text{vir}$ is included to bracket the uncertainty from above. 

The resulting $\epsilon(z)$ for these three
cases are shown in Fig.~\ref{fig:epsilon} for the cosmology considered in this work. They have a very similar $z$-dependence and differ mainly in their normalization,
with $k_\sigma\equiv \epsilon(0)^{-1/2}=0.70, 0.36, 0.25\, h/$Mpc, respectively. We find that the $z$-dependence can be fit by a broken power law of the form
$\epsilon(z)=k_\sigma^{-2}D(z)^\alpha g(z)/g(0)$ with $g(z)=1/(1 +(D(z_c)/D(z))^{p\,\Delta\alpha})^{1/p}$ with $p=5$, such that $\epsilon(z)\to k_\sigma^{-2}D(z)^\alpha$ for $z\ll z_c$
and $\epsilon(z)\propto D(z)^{\alpha+\Delta\alpha}$ for $z\gg z_c$. A fit for our fiducial case gives $\alpha=3.23$,
$\alpha+\Delta\alpha=4.53$ and $z_c=1.52$. The parameters for the other two cases are very similar. The broken power-law fit as well
as the single power-law approximation for $z\ll z_c$ are also shown as dashed (dotted) lines in Fig.~\ref{fig:epsilon}.

\begin{figure*}[t]
  \begin{center}
  \includegraphics[width=\textwidth]{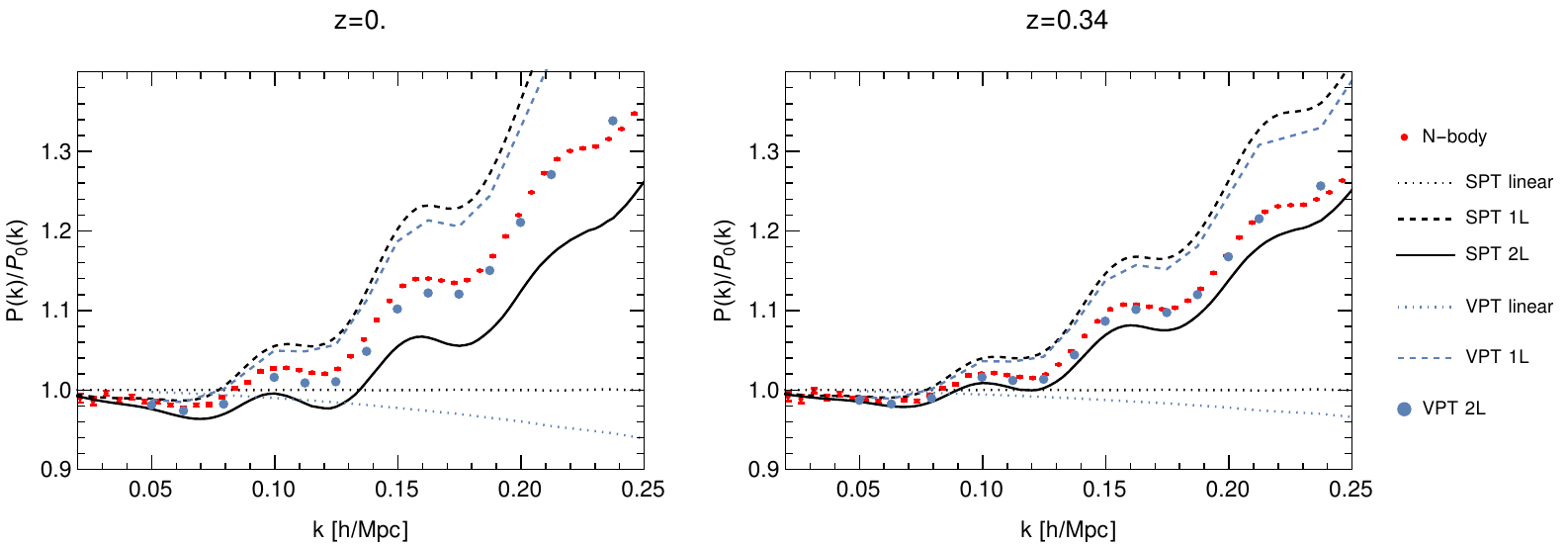}
  \end{center}
  \caption{\label{fig:loops}
  \vpt{} matter power spectrum in linear (dotted), one-loop (dashed) and two-loop (circles) approximation at $z=0$ (left) and $z=0.34$ (right).
  $N$-body data are shown in red, and SPT results in black for comparison. Here we use our fiducial setup with $c_\text{max}=2$ and $\epsilon(z)$ from halos cut at $1.5r_\text{vir}$,  see Figs.\,\ref{fig:eps}-\ref{fig:cutoff} for a discussion of the (in-)dependence on these settings. \vpt{} residuals from simulations at $z=0$ and $z=0.34$ are consistent with a growth-factor scaling $[D(z)]^8$, indicating evidence for the need of three-loop corrections.
  }
\end{figure*}

%===========================================================
\section{Matter power spectrum in \vpt{}}
\label{sec:pk}
%===========================================================

Our result for the matter power spectrum in linear, one-loop and two-loop
approximation within \vpt{} is shown in Fig.\,\ref{fig:loops}, at $z=0$ (left)
and $z=0.34$ (right), respectively. Here we use our fiducial choice, being the truncation
$c_\text{max}=2$ (see Table~\ref{tab:modes}) and average dispersion from halos, see Eq.\,\eqref{eq:halo}, cut at $1.5r_\text{vir}$ (middle line in Fig.\,\ref{fig:epsilon}),
with $k_\sigma=0.36h/$Mpc  at $z=0$ (for comparison $k_{\rm nl}=0.27h/$Mpc) and $k_\sigma=0.47h/$Mpc  at $z=0.34$ ($k_{\rm nl}=0.37h/$Mpc). We discuss the (in-)dependence on these choices below.

We see that the linear \vpt{} result already differs from SPT (dotted lines in Fig.\,\ref{fig:loops}), showing the onset of suppression when $k$ approaches 
the dispersion scale $k_\sigma$. This suppression is partly compensated at one-loop (dashed lines), which we associate with a less negative contribution
from $P_{13}$ (second line in Eq.\,\ref{eq:P1L2L}) due to the impact of screening on the $F_3$ kernel. This effect is much more pronounced for $P_{15}$
at two-loop order (fifth line in Eq.\,\ref{eq:P1L2L}), where the $F_5$ kernel shows a spurious UV dependence in SPT that is not present in \vpt{}.
This makes the two-loop correction smaller in magnitude than in SPT, {\it i.e.} less negative, leading to a larger total result for the power spectrum (full circles vs. solid
lines in Fig.\,\ref{fig:loops}). As is well known, within SPT, the spurious UV sensitivity from treating small-scale modes as obeying the dynamics of an ideal pressureless fluid translates into a significant dependence of the two-loop result on the UV cutoff used in the loop integration. In contrast, the \vpt{} result is independent of the cutoff
(see Appendix\,\ref{app:cutoff} for more details on this).

The two-loop power spectrum in \vpt{} matches the $N$-body result very well for $z=0.34$, with less than $\sim 1\%$ deviation
up to $\sim 0.22h/$Mpc. We stress that for the \vpt{} result, {\it no} free parameters were adjusted.  We also checked that the agreement with $N$-body data persists at higher redshift.
At $z=0$, the two-loop \vpt{} prediction falls slightly below
the $N$-body result at $k\lesssim 0.2h/$Mpc, with a maximal deviation of $2\%$ at $k\sim 0.15h/$Mpc. This small deviation is fully compatible with missing contributions
from even higher loop orders. In particular, the residual deviations between two-loop and $N$-body results are expected to increase with $[D(z)]^8$, implying a relative
increase of order $(D(0)/D(0.34))^8\sim 3.7$ between $z=0.34$ and $z=0$. We checked that the \vpt{} result is consistent with this expectation.

%----------------------------------------------------------
\subsection{Dependence on average velocity dispersion}\label{sec:ksigmadependence}
%----------------------------------------------------------

\begin{figure*}[t]
  \begin{center}
  \includegraphics[width=\textwidth]{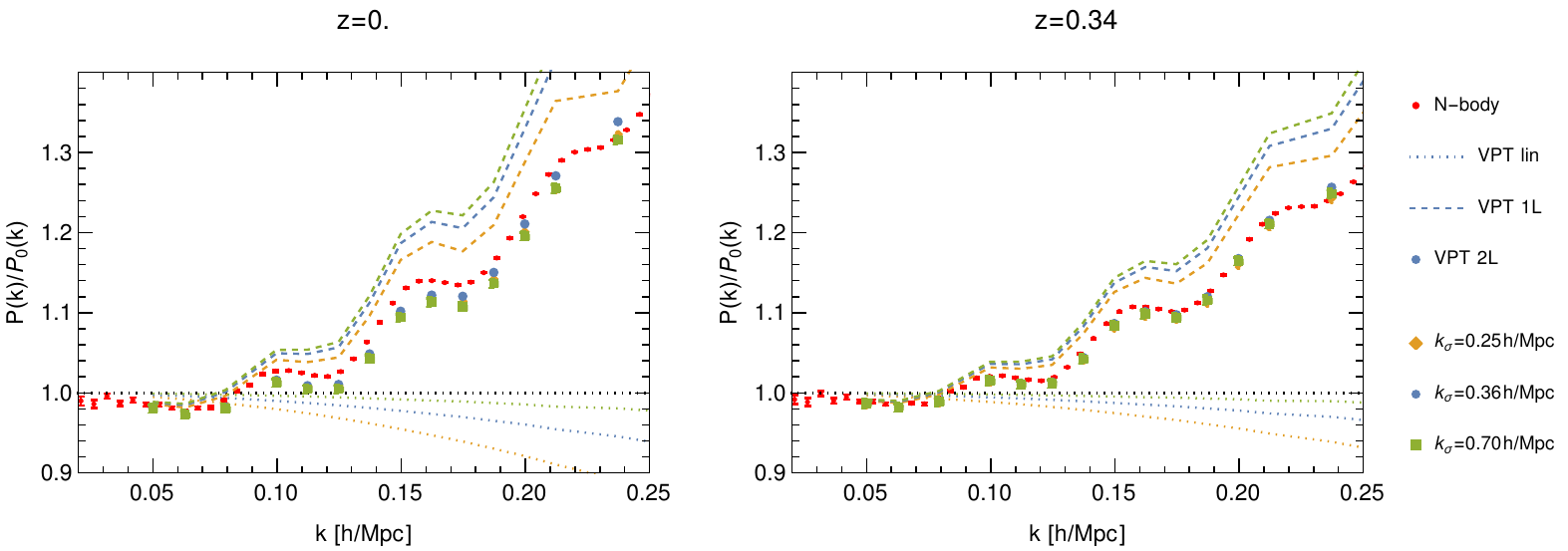}
  \end{center}
  \caption{\label{fig:eps}
  Dependence of the matter power spectrum on the average dispersion $\epsilon(z)$ in linear (dotted), one-loop (dashed) and two-loop (symbols) approximation. The different colors correspond to results obtained from Eq.\,\eqref{eq:halo} for the three cases for $\epsilon(z)$ shown in Fig.\,\ref{fig:epsilon},  that conservatively bracket the viable range. The dispersion scale at $z=0$ is $k_\sigma=0.70, 0.36, 0.25\, h/$Mpc, respectively. At two-loop the results lie almost on top of each other, especially for $z=0.34$. The non-monotonic dependence on $k_\sigma$ for the two-loop case is evident at $z=0$ (see Fig.~\ref{fig:vpteffksigdep} for more details on this).
  }
\end{figure*}

\begin{figure*}[t]
  \begin{center}
  \includegraphics[width=\textwidth]{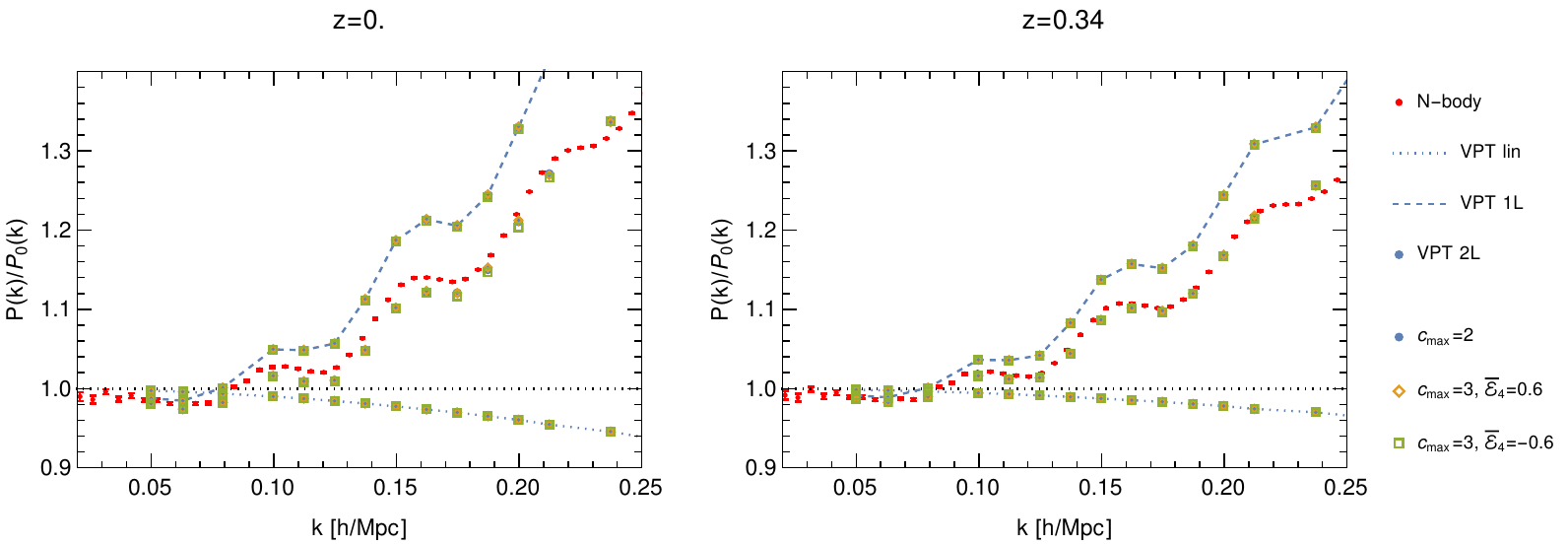}
  \end{center}
  \caption{\label{fig:cmax}
  Dependence of the matter power spectrum on the truncation order $c_\text{max}$ of the cumulant hierarchy, in linear, one-loop and two-loop approximation. The differences between $c_\text{max}=2$ (blue) and $c_\text{max}=3$ are very small, with all results lying almost on top of each other.
  For $c_\text{max}=3$, two results for $\bar{\cal E}_4=\pm 0.6$ are shown (diamond and square symbols) for each loop order, respectively.
  }
\end{figure*}

Let us now discuss the impact of various settings and approximations within the \vpt{} setup.
Fig.\,\ref{fig:eps} shows the dependence on the background dispersion $\epsilon(z)$, using the three choices displayed in Fig.\,\ref{fig:epsilon},
that conservatively bracket physically viable values following from Eq.\,\eqref{eq:halo}.
Consequently, these cases span a wide range for the dispersion scale, with values at $z=0$ given by $k_\sigma=0.70, 0.36, 0.25\, h/$Mpc, respectively.
As expected, the linear \vpt{} result (dotted) is the more suppressed the larger $k_\sigma$. The same holds at one-loop (dashed). However, remarkably, the two-loop
result is almost independent of $k_\sigma$ (symbols). This occurs because the two-loop correction becomes smaller in magnitude for larger $k_\sigma$, due to
stronger UV screening of $P_{15}$. Thus, it is less negative and therefore counteracts the suppression of the linear and one-loop results. 

This cancellation has the
interesting consequence that the \vpt{} prediction for the matter power spectrum is almost insensitive to $k_\sigma$ at the percent level, even for the wide range covered by the three
choices used in Fig.\,\ref{fig:epsilon} that differ almost by a factor of three. While this particular feature may be a coincidence that does not occur for other summary statistics, it
is nevertheless relevant due to the important role of the matter power spectrum. Moreover, as we discuss in more detail below, the competition between linear suppression and screening of $P_{15}$ leads to a non-monotonic variation of the power
spectrum when changing $k_\sigma$. Thus, within the range $k\lesssim 0.2h/$Mpc, the \vpt{} two-loop result attains a maximum at a certain value of $k_\sigma$. Indeed, it can be seen
in Fig.\,\ref{fig:eps} that the two-loop result obtained for $k_\sigma=0.36\, h/$Mpc (blue circles) is larger than for both $0.70$ and $0.25\, h/$Mpc. This implies that even when allowing for a wide variation of $k_\sigma$,
the resulting power spectrum cannot be adjusted arbitrarily, but rather varies only within a small band. This can also be seen clearly in Fig.\,\ref{fig:Pdd2L}, see also Sec.\,\ref{sec:vpt} where we explore this in more detail.

%----------------------------------------------------------
\subsection{Dependence on truncation}
%----------------------------------------------------------

To assess the impact of truncating the Vlasov hierarchy at a certain cumulant order, we compare the fiducial result obtained for $c_\text{max}=2$
with that obtained when including also all scalar and vector modes of the third cumulant, as well as all non-linear interactions among modes up
to this order resulting from the Vlasov-Poisson equations (the $c_\text{max}=3$ scheme in Table~\ref{tab:modes}).  The perturbation equations in
this truncation depends on the average value of the fourth cumulant, and we consider two choices $\bar{\cal E}_4=\pm 0.6$.
These values are consistent with constraints from requiring stability of the coupled set of perturbation equations for
scalar and vector (and higher) modes~\cite{cumPT,Garny:2025ovs}, being $-6\leq \bar{\cal E}_4\leq 2$.

The \vpt{} results for $c_\text{max}=3$ shown in Fig.\,\ref{fig:cmax} are almost indistinguishable from those for $c_\text{max}=2$.
This is in line with previous findings for scale-free models~\cite{Garny:2025ovs}. While higher cumulants influence \vpt{} kernels on
scales sufficiently far above $k_\sigma$, this region gives a negligible contribution to the integrated loop corrections since the
kernels are highly suppressed on these UV scales. Thus, the independence on the truncation order beyond $c_\text{max}=2$ is also
a consequence of UV screening captured already by $c_\text{max}=2$ within \vpt{}.

%===========================================================
\section{Fast \vpt{}}
\label{sec:vpt}
%===========================================================

\begin{figure*}[t]
  \begin{center}
  \includegraphics[width=\textwidth]{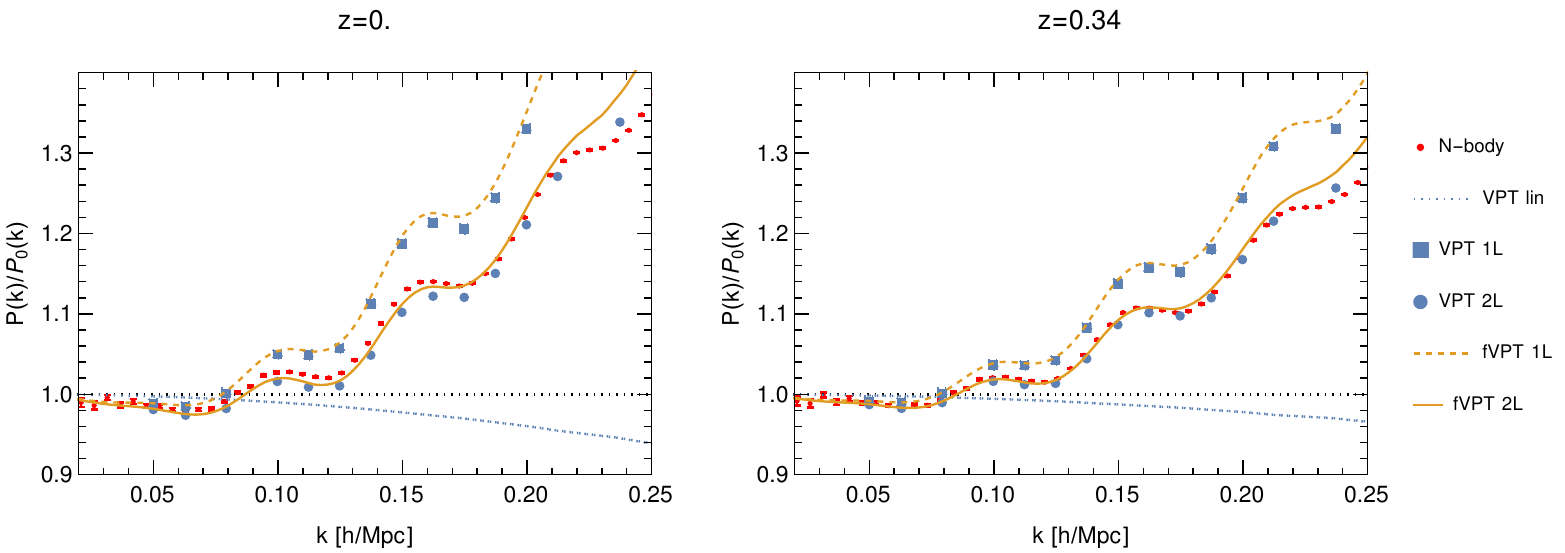}
  \end{center}
  \caption{\label{fig:vptvpteff}
  Comparison of full \vpt{} results (squares and circles at one- and two-loop) with \fastvpt{} (dashed and solid lines at one- and two-loop, using kernels from Eq.\,\ref{eq:fastvpt}), for the same underlying average dispersion $\epsilon(z)$ as in Fig.\,\ref{fig:loops}.
  The linear results (dotted) are identical in full and fast \vpt{} by construction. Both one- and two-loop agree in full and fast \vpt{} at the percent level.
  }
\end{figure*}

\begin{figure*}[t]
  \begin{center}
  \includegraphics[width=\textwidth]{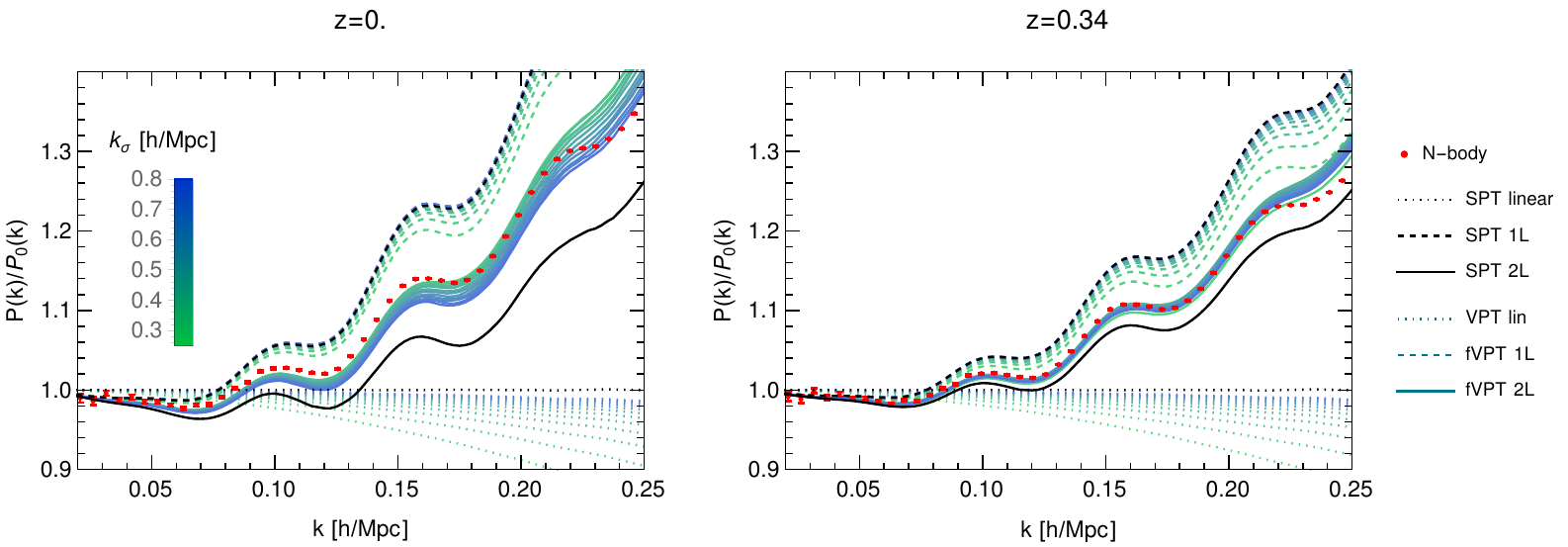}
  \end{center}
  \caption{\label{fig:vpteff}
  Dependence of \vpt{} linear (dotted), one- (dashed) and two-loop (solid) result on the dispersion scale $k_\sigma\equiv\epsilon(0)^{-1/2}=0.25,0.3,\dots,0.8h/$Mpc, for $\epsilon(z)\propto D(z)^\alpha$ with $\alpha=3.3$.
  The two-loop result is almost insensitive to $k_\sigma$ despite the variation over a very large range, due to a cancellation of the
  impact of $k_\sigma$ on the linear and (two-)loop contributions. This feature, observed before in full \vpt{} (see Fig.\,\ref{fig:eps}),
  can thus be well reproduced in the \fastvpt{} approximation employed here, using Eqs.\,(\ref{eq:fastvpt}, \ref{eq:F1G1analytic}).
  }
\end{figure*}

\begin{figure*}[t]
  \begin{center}
  \includegraphics[width=\textwidth]{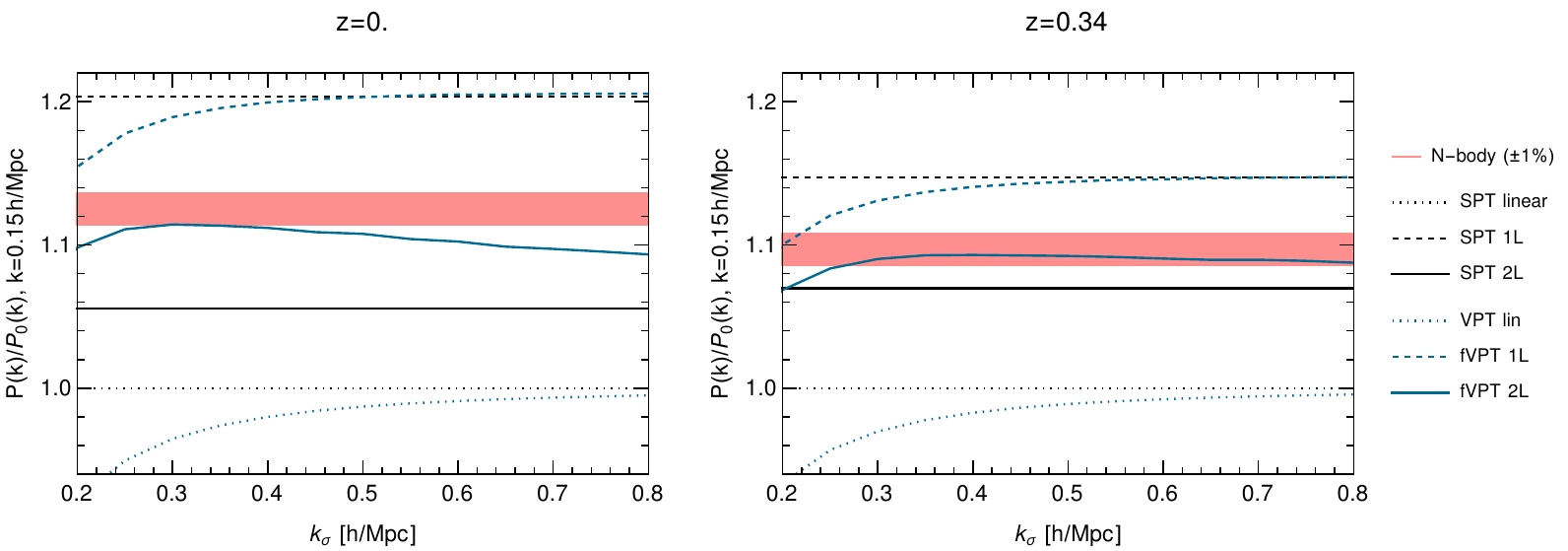}
  \end{center}
  \caption{\label{fig:vpteffksigdep}
  Dependence of the \vpt{} linear (blue dotted), one-loop (blue dashed) and two-loop (blue solid) matter power spectrum on the average dispersion scale $k_\sigma$, for
  fixed $k=0.15h/$Mpc and $\alpha=3.3$ using \fastvpt{}. For comparison, the (constant) $N$-body (red) and SPT (black) results at $k=0.15h/$Mpc are shown as horizontal
  lines. While linear and one-loop result increase monotonically, the two-loop features a maximum and varies only very mildly despite changing $k_\sigma$ over a wide range.
  The non-monotonic behaviour can be traced back to a cancellation between the impact of $k_\sigma$ on the two-loop contribution (dominantly $P_{15}$) and on the linear \vpt{} result.
  }
\end{figure*}

\begin{figure*}[t]
  \begin{center}
  \includegraphics[width=\textwidth]{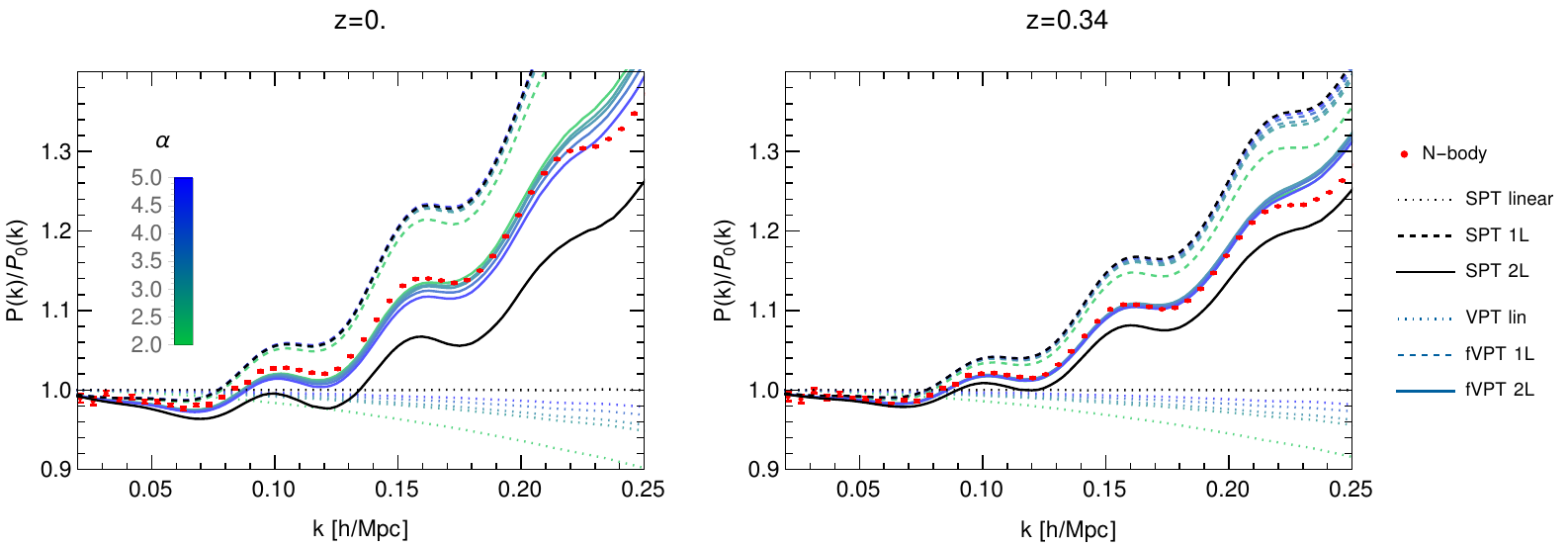}
  \end{center}
  \caption{\label{fig:vpteffalpha}
  Similar as Fig.\,\ref{fig:vpteffksigdep}, but varying the power-law exponent $\alpha=2,3,3.3,4,5$ of the average dispersion $\epsilon(z)\propto D(z)^\alpha$ while
  keeping $k_\sigma\equiv\epsilon(0)^{-1/2}=0.45h/$Mpc fixed. We observe a very mild dependence of the two-loop result, despite a significant variation of $\alpha$.
  This further confirms that the two-loop \vpt{} result is a robust prediction of collisionless dynamics.
  }
\end{figure*}

Above we found that the \vpt{} result for the matter power spectrum yields a genuine prediction for
gravitational clustering of collisionless matter, being robust to assumptions on truncation of the hierarchy as well as on the value of the average velocity dispersion.
Since the extended set of perturbation modes and their non-linear interactions makes \vpt{} numerically more expensive than SPT,
it is desirable to search for an approximate treatment of velocity dispersion effects that allows for quick numerical evaluation.
Here we propose such a scheme, that we dub ``\fastvpt{}'', and then use it to investigate the dependence of the matter power spectrum
on the average velocity dispersion in more detail.

%----------------------------------------------------------
\subsection{Approximate \vpt{} kernels}
%----------------------------------------------------------

The starting point for \fastvpt{} is the observation that (i) already linear evolution in \vpt{} is
non-trivial and captures suppression of modes with $k\gtrsim k_\sigma$, and (ii) for $\Lambda$CDM cosmologies the quantitative differences between one- and especially two-loop integrals in \vpt{} and SPT arise mainly from the integration region where \vpt{} kernels are only moderately suppressed compared to the SPT case.
Thus, it is most important to capture the onset of suppression in linear and non-linear kernels when approaching the dispersion scale $k_\sigma$.
While the kernels in the deeply screened region $p,q\gg k_\sigma$ are sensitive to the full and precise \vpt{} description, and require the
inclusion of higher and higher cumulants when approaching smaller and smaller scales, this region does not yield a sizable contribution to
the loop integrals for the matter power spectrum evaluated on weakly non-linear scales.  

This motivates the following ansatz for non-linear kernels of the density and velocity fields,
\bea\label{eq:fastvpt}
  F_n^\text{\fastvpt{}}({\bm k}_1,\dots,{\bm k}_n;z) &=& F_n^\text{SPT}({\bm k}_1,\dots,{\bm k}_n)\nn\\
  &\times&  F_1^\text{\vpt{}}(k_1,z)\cdots F_1^\text{\vpt{}}(k_n,z)\,,\nonumber\\
  G_n^\text{\fastvpt{}}({\bm k}_1,\dots,{\bm k}_n;z) &=& G_n^\text{SPT}({\bm k}_1,\dots,{\bm k}_n)\nn\\
  &\times&   G_1^\text{\vpt{}}(k_1,z)\cdots G_1^\text{\vpt{}}(k_n,z)\,,\nonumber\\
\eea
where $F_n^\text{SPT}, G_n^\text{SPT}$ are the standard SPT kernels, and $F_1^\text{\vpt{}}(k,z), G_1^\text{\vpt{}}(k,z)$
the linear \vpt{} kernels for the density field $\delta$ and the normalized velocity divergence $\theta$, respectively.
Both approach unity on large scales, such that \vpt{} approaches SPT when all arguments are small, as expected.
The suppression of $F_1^\text{\vpt{}}(k,z)$ and $G_1^\text{\vpt{}}(k,z)$ when approaching $k_\sigma$ approximately captures the onset of UV
screening predicted in full \vpt{}. Finally, by construction, the linear kernels in full and fast \vpt{} agree exactly. In addition, the ansatz in Eq.~(\ref{eq:fastvpt}) satisfies the required symmetries, i.e. constraints from momentum conservation which requires $F_n \propto k^2$ when ${\bm k} \equiv \sum_i {\bm k}_i \to 0$, and Galilean invariance which requires that in the squeezed limit ${\bm p}\to 0$, 
\be
F_{n+1}({\bm p},{\bm k}_1,\dots,{\bm k}_n;z) \to {1\over {n+1}} \frac{ {\bm k}\cdot {\bm p}}{p^2}\ F_{n}({\bm k}_1,\dots,{\bm k}_n;z)\;,
\label{squeezed}
\ee
which holds in full \vpt{}~\cite{cumPT2} and SPT~\cite{Sugiyama:2013pwa}, and in \fastvpt{} as a result of $F_1^\text{\vpt{}}(p,z) \to 1$ as $p\to 0$ and the corresponding SPT result.

In general, for a given truncation order and choice of average cumulants, $F_1^\text{\vpt{}}(k,z)$ and $G_1^\text{\vpt{}}(k,z)$ can be computed numerically by solving Eq.\,\eqref{eq:psieom}
for the coupled set of cumulant perturbation modes in linear approximation. However, since Eq.\,\eqref{eq:fastvpt} is expected to be
a reasonable approximation only as long as wavenumbers are not too far above $k_\sigma$, using our fiducial choice $c_\text{max}=2$ for computing
the linear kernels is sufficient, since the differences between second and higher cumulant truncations are
small on those scales~\cite{cumPT,Garny:2025ovs}.
When adopting in addition the power-law parameterization Eq.\,\eqref{eq:powerlaw} for the average dispersion, as well as the common EdS approximation, analytical expressions for the linear
kernels in terms of generalized hypergeometric functions ${}_1F_2$ exist~\cite{cumPT},
\bea\label{eq:F1G1analytic}
  F_1^\text{\vpt{}}(k,z) &=& {}_1F_2\left(\frac{4+\alpha}{3\alpha};1+\frac{2}{\alpha},1+\frac{5}{2\alpha};\frac{-3k^2\epsilon(z)}{\alpha^2}\right)\,,\nn\\
  G_1^\text{\vpt{}}(k,z) &=& F_1^\text{\vpt{}}(k,z) - \frac{2(4+\alpha)k^2\epsilon(z)}{(2+\alpha)(5+2\alpha)}\nn\\
  &\times&  {}_1F_2\left(\frac{4+4\alpha}{3\alpha};2+\frac{2}{\alpha},2+\frac{5}{2\alpha};\frac{-3k^2\epsilon(z)}{\alpha^2}\right)\,.\nn\\
\eea
The \fastvpt{} kernels in Eq.\,\eqref{eq:fastvpt} can be easily evaluated, with the same numerical cost as in SPT. When computing the matter density
power spectrum, it is equivalently possible to rescale the input power spectrum as $P_0(k)\mapsto F_1^\text{\vpt{}}(k,z)^2P_0(k)$, making \fastvpt{}
compatible with {\em e.g.}~loop evaluations based on fast Fourier transformation methods~\cite{Simonovic:2017mhp}.

As a validation check, we compare full and fast \vpt{} results in Fig.\,\ref{fig:vptvpteff}, for identical choice of $\epsilon(z)$ as in Fig.\,\ref{fig:loops}.
We find that differences between full and fast \vpt{} are below the percent level both at $z=0$ and $z=0.34$, and are comparable to or smaller than the
sensitivity to $\epsilon(z)$, see Fig.\,\ref{fig:eps}. For this comparison, we use numerically computed linear \vpt{} kernels that are identical to those for the full \vpt{} setup. 
However, we find that using the analytical power-law result instead, with exponent $\alpha$ matched to $\epsilon(z)$ at low redshift, yields very similar results.

Thus, we find that the \fastvpt{} approximation is accurate at the (sub-)percent level (particularly given that $k_\sigma=\epsilon^{-1/2}$ is an adjustable parameter) and a useful approach, greatly facilitating numerical evaluation and implementation.

%----------------------------------------------------------
\subsection{Cancellation of the dependence on $k_\sigma$}\label{sec:ksigmadependencefvpt}
%----------------------------------------------------------

In this section we return to the finding from Sec.\,\ref{sec:ksigmadependence} concerning the dependence of the \vpt{} matter power
spectrum on the average dispersion $\epsilon(z)$. In \fastvpt{} with analytical linear kernels, the matter power spectrum depends
on the average dispersion via the two parameters $k_\sigma=\epsilon(0)^{-1/2}$ and $\alpha=d\ln\epsilon/d\ln D(z)$, see Eq.\,\eqref{eq:powerlaw}.
In Fig.\,\ref{fig:vpteff}, we show the variation of the linear, one-loop and two-loop matter power spectrum when varying $k_\sigma$, while
fixing $\alpha=3.3$. As noted before, the two-loop result is almost insensitive due to a cancellation of the impact of $k_\sigma$ on the linear spectrum
and on $P_{15}$. This feature is also captured by \fastvpt{}. Remarkably, the uncertainty band of the two-loop result at $z=0.34$ is only at the
(sub-)percent level for $k\lesssim 0.2h/$Mpc, even though $k_\sigma$ is varied over a very large range $0.25-0.8h/$Mpc. At $z=0$ the relative variation
slightly increases to about $2-4\%$ at $k\sim 0.15-0.2\, h/$Mpc.

To illustrate the non-monotonic behaviour of the two-loop power spectrum when varying $k_\sigma$, we show the dependence on this scale for a fixed value of $k=0.15h/$Mpc
in Fig.\,\ref{fig:vpteffksigdep}. While both the linear and one-loop \vpt{} results increase monotonically with $k_\sigma$, the two-loop features a maximum and a very
flat plateau-like behaviour. All \vpt{} results approach the corresponding SPT limit for $k_\sigma\to\infty$, since the kernels within the relevant integration domain
approach the SPT limit when $k/k_\sigma, p/k_\sigma, q/k_\sigma\to 0$. Note that this would be different for scaling models for which the one- and two-loop integrals are UV divergent in SPT, since
they feature relevant contributions to the loop integral within \vpt{} for $p, q\sim k_\sigma$ no matter how large $k_\sigma$ is. Nevertheless, even for $\Lambda$CDM initial spectra, the fact that the two-loop result approaches the
SPT limit only very slowly when increasing $k_\sigma$ (see blue and black solid lines in Fig.\,\ref{fig:vpteffksigdep}) is related to the large, spurious UV sensitivity of $P_{15}$ in SPT. In contrast, the \vpt{}
result is regulated by the UV screening captured by \vpt{}. 

Finally, we also assess the sensitivity of the matter power spectrum on $\alpha$, which characterizes the redshift-dependence of the average dispersion.
In Fig.\,\ref{fig:vpteffalpha} we show results for the wide range $\alpha=2,\dots,5$. Again, the dependence of the two-loop result is very mild. This further supports our
finding that the \vpt{} prediction is robust despite allowing for large variations in the average dispersion.

%----------------------------------------------------------
\subsection{Relation to EFT approach}
%----------------------------------------------------------

While \vpt{} can be viewed as a (conceptually straightforward) perturbative
solution of the Vlasov-Poisson system for collisionless dynamics, the EFT approach~\cite{BauNicSen1207}
is completely agnostic about the dynamics on scales for which the pressureless perfect
fluid approximaton breaks down. The price for this ignorance is the appearance of
free parameters, that are unconstrained and need to be fitted to simulations
or data.

The EFT expansion yields a priori an infinite set of correction terms, with associated
free parameters, that can be classified according the order in perturbation theory
and in a derivative expansion in powers of $k/k_\text{nl}$. The leading correction term in
the derivative expansion, and at one-loop order, is $\Delta P = c_s^2k^2P_0(k)$ with a
free parameter $c_s^2$. In addition, at one-loop the leading noise term yields another
additive correction $\propto k^4$, being suppressed on small scales due to momentum conservation.
At two-loop order, in principle all EFT terms consistent with momentum conservation constraints
that renormalize up to fifth order kernels need to be added. Even when disregarding noise terms
and restricting to the leading terms in the derivative expansion, there are four terms for renormalizing the
fourth order kernel~\cite{BalMerMir1505}, and (at least) $13$ terms at fifth order~\cite{Schmidt:2020tao}, leading to $1+4+13=18$ free parameters (for each considered redshift).
In practice, not all of these are included,  {\it e.g.}~\cite{BalMerZal1512} use a two parameter EFT model (at a given redshift) at two-loop order.

The \vpt{} approach differs from the EFT by being informed about the collisionless dynamics. Thus, a priori, \vpt{} has no free parameters.
Moreover, it does not employ a derivative expansion, but resums corrections of higher powers in $k/k_\sigma$, being instrumental for
capturing UV screening. In practice, the average dispersion $\epsilon(z)$ is taken as non-perturbative input in \vpt{}. However, we find
that the dependence on the parameters $k_\sigma$ and $\alpha$ associated to $\epsilon(z)$ (see Eq.\,\ref{eq:powerlaw}) is highly non-trivial, and allows
only for a very small variation in the two-loop power spectrum. In this sense $k_\sigma$ and $\alpha$ cannot be regarded on the same footing as EFT parameters.
Furthermore, the average dispersion $\epsilon(z)$  has a clear physical interpretation, in contrast to EFT corrections that need to absorb the errors made by SPT.
Moreover, as demonstrated in~\cite{cumPT2,Garny:2025ovs}, \vpt{} is able to
predict also other summary statistics such as the bispectrum or velocity power spectra, for which additional free parameters would be
needed in an EFT treatment.

Nevertheless, one may also consider the possibility of adding EFT corrections to \vpt{}, to account for effects on the power spectrum beyond
collisionless clustering, or mimick the leading contribution from missing higher-order corrections at two-loop order. Since the \vpt{} result is
already close to $N$-body data, we expectedly find that when adding a $c_s^2$-correction to \vpt{}, the fit to $N$-body data improves significantly
only at $z=0$ where \vpt{} is slightly below the $N$-body result. As compared to SPT, the value of $c_s^2$ at $z=0$ is a factor of $\sim 3$ smaller.
Thus, \vpt{} can allow for putting tighter priors on EFT corrections, since it yields predictions that are closer to data than SPT.
We leave a more detailed analysis of this possibility to future work.

%===========================================================
\section{Conclusions}
\label{sec:conclusions}
%===========================================================

We apply the framework of \vpt{} to $\Lambda$CDM cosmology and present results for
the matter power spectrum at up to two-loop order. Our main findings are:

\begin{itemize}

\item[i)] The impact of the truncation of the cumulant hierarchy is well below the percent level on weakly non-linear scales. Including scalar and vector modes of the velocity dispersion tensor, as well as the average dispersion, captures the physics that is
missed in SPT while third-cumulant perturbations play only a negligible role quantitatively. This can be understood in terms of the physically expected screening that is captured in \vpt{} but not in SPT, making the result of loop computations largely insensitive to UV scales. The \vpt{} two-loop result is also independent of the UV cutoff, in contrast to SPT.

\item[ii)] When using the average velocity dispersion estimated from a halo model as input, the \vpt{} result agrees at the (sub-)percent level with $N$-body data up to $k\sim 0.22h/$Mpc without any free parameters at $z=0.34$. At $z=0$, the \vpt{} prediction is
slightly below the $N$-body result, by about $\sim 2\%$ at $k=0.15h/$Mpc. This can be attributed to missing higher loop corrections, as indicated by the scaling with redshift of the residuals.

\item[iii)] We find that the two-loop \vpt{} result for the matter power spectrum is remarkably insensitive to the value of the average dispersion, and thus the associated dispersion scale $k_\sigma$. Even when varying $k_\sigma$ by a factor of $\sim 3$, the two-loop
result at $k\lesssim 0.2h/$Mpc changes only by $\lesssim 1\%$ ($\lesssim 4\%$) at $z=0.34$ ($z=0$).  This can be understood by a cancellation of the $k_\sigma$-dependence of linear and loop contributions occuring for the matter power spectrum. We also find  insensitivity to the redshift-dependence of the average dispersion at a similar level. 

\item[iv)] We propose and validate an approximate treatment of the impact of velocity dispersion on the density and velocity fields, dubbed \fastvpt{}, that can be easily included in existing perturbation theory codes. The effort of numerical evaluation for \fastvpt{} is identical to SPT. We show that it captures
the dominant effects from velocity dispersion on the matter power spectrum, including in particular UV screening, and agrees with full \vpt{} at the percent level on weakly non-linear scales.

\end{itemize}

Altogether, we find that the \vpt{} two-loop result is a robust prediction for gravitational clustering of collisionless matter in $\Lambda$CDM cosmologies at weakly non-linear scales.
It is sufficient to achieve percent accuracy for the matter power spectrum at $k\lesssim 0.2h/$Mpc and $z\lesssim 0.3$, while the dependence on uncertainties in the knowledge of the average velocity dispersion generated
from small-scale shell crossing is remarkably small. We expect a somewhat stronger dependence for power spectra sensitive to the velocity field, such as in redshift space, opening the possibility of inferring the
scale $k_\sigma$, while maintaining predictivity even at two-loop order. We leave a detailed analysis of this topic to future work.

\vspace*{2em}
\acknowledgments

We acknowledge support by the Excellence Cluster ORIGINS, which is funded by the Deutsche Forschungsgemeinschaft (DFG, German Research
Foundation) under Germany's Excellence Strategy - EXC-2094 - 390783311 and by Deutscher Akademischer Austauschdienst (DAAD).

\appendix

%===========================================================
\section{Cutoff independence}
\label{app:cutoff}
%===========================================================

\begin{figure*}[t]
  \begin{center}
  \includegraphics[width=\textwidth]{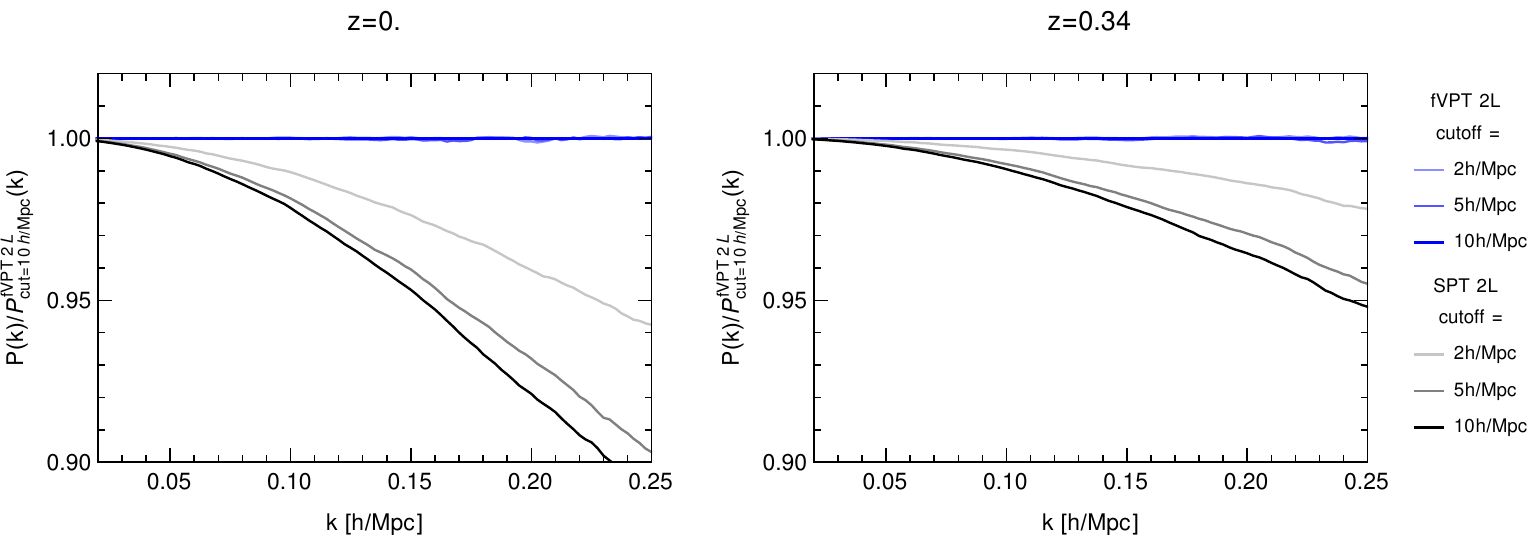}
  \end{center}
  \caption{\label{fig:cutoff}
  (In-)dependence of two-loop power spectrum on the UV cutoff in \vpt{} (using \fastvpt{} with $k_\sigma=0.5h/$Mpc, $\alpha=3.3$ for illustration) and SPT.
  Note that all blue lines for VPT lie almost on top of each other, being a consequence of UV screening captured by \vpt{}, while the SPT result is cutoff-dependent.
  All results are normalized to the VPT two-loop spectrum for a cutoff of $10h$/Mpc.
  }
\end{figure*}

Here we demonstrate the independence of the \vpt{} two-loop result for the power spectrum
on the UV cutoff used for the loop integration in Fig.\,\ref{fig:cutoff}. We also show
the corresponding SPT results for comparison, featuring the well-known unphysical cutoff-dependence, as a result of the
ideal pressure-less fluid approximation not capturing the UV screening arising from velocity dispersion and higher cumulants.

% Create the reference section using BibTeX:
%\bibliography{refs}
%merlin.mbs apsrev4-1.bst 2010-07-25 4.21a (PWD, AO, DPC) hacked
%Control: key (0)
%Control: author (8) initials jnrlst
%Control: editor formatted (1) identically to author
%Control: production of article title (-1) disabled
%Control: page (0) single
%Control: year (1) truncated
%Control: production of eprint (0) enabled
%

\end{document}